\documentclass[aip, apl, reprint]{revtex4-1}
\usepackage{amsmath}
\usepackage{amssymb}
\usepackage{graphicx}
\usepackage{color}

\begin{document}

\title{Observation of suppressed terahertz absorption in photoexcited graphene}
\author{A. J. Frenzel}
\affiliation{Department of Physics, Massachusetts Institute of Technology, Cambridge, Massachusetts 02139, USA.}
\affiliation{Department of Physics, Harvard University, Cambridge, Massachusetts 02138, USA.}
\author{C. H. Lui}
\affiliation{Department of Physics, Massachusetts Institute of Technology, Cambridge, Massachusetts 02139, USA.}
\author{W. Fang}
\affiliation{Department of Electrical Engineering and Computer Science, Massachusetts Institute of Technology, Cambridge, Massachusetts 02139, USA.}
\author{N. L. Nair}
\affiliation{Department of Physics, Massachusetts Institute of Technology, Cambridge, Massachusetts 02139, USA.}
\author{P. K. Herring}
\affiliation{Department of Physics, Massachusetts Institute of Technology, Cambridge, Massachusetts 02139, USA.}
\affiliation{Department of Physics, Harvard University, Cambridge, Massachusetts 02138, USA.}
\author{P. Jarillo-Herrero}
\affiliation{Department of Physics, Massachusetts Institute of Technology, Cambridge, Massachusetts 02139, USA.}
\author{J. Kong}
\affiliation{Department of Electrical Engineering and Computer Science, Massachusetts Institute of Technology, Cambridge, Massachusetts 02139, USA.}
\author{N. Gedik}
\email{gedik@mit.edu}
\affiliation{Department of Physics, Massachusetts Institute of Technology, Cambridge, Massachusetts 02139, USA.}

\date{\today}

\begin{abstract}
When light is absorbed by a semiconductor, photoexcited charge carriers enhance the absorption of far-infrared radiation due to intraband transitions.   We observe the opposite behavior in monolayer graphene, a zero-gap semiconductor with linear dispersion.  By using time domain terahertz (THz) spectroscopy in conjunction with optical pump excitation, we observe a reduced absorption of THz radiation in photoexcited graphene.  The measured spectral shape of the differential optical conductivity exhibits non-Drude behavior.  We discuss several possible mechanisms that contribute to the observed low-frequency non-equilibrium optical response of graphene. 
\end{abstract}

\maketitle

Two-dimensional graphene is characterized by its distinctive Dirac electronic structure and the associated remarkable optical properties, specifically, a strong and broadband optical absorption from far-infrared to ultraviolet wavelengths.\cite{Geim2007,Bonaccorso2010,Mak2012} The unique absorption spectrum of graphene, together with the great tunability of its Fermi level by electrical gating, has made it a promising material for next-generation optoelectronic applications in the far-infrared regime. For instance, recent research has demonstrated graphene to be an efficient terahertz (THz) modulator\cite{Horng2011, Ren2012, Maeng2012} and a possible gain medium for a THz laser.\cite{Otsuji2012}  Optical pump excitation has been shown to effectively modulate the THz response of graphene.  In particular, recent studies report enhanced absorption of THz radiation in optically pumped graphene.\cite{George2008,Choi2009,Strait2011}  In these works, the observation was understood by considering Drude absorption with a fixed scattering rate, which typically increases with the population of photoexcited carriers.  However, the small Fermi energy of charge carriers in graphene suggests that non-Drude behavior is likely to occur at the high transient temperatures encountered in pump-probe experiments, and the coupling of carriers to hot phonons may alter their scattering rates. These effects on the low-frequency optical response of graphene under non-equilibrium conditions have not yet been thoroughly investigated.

In this paper, we present optical-pump THz-probe\cite{Beard2001,Ulbricht2011}   measurements of the ultrafast far-infrared response of large-area monolayer graphene grown by chemical vapor deposition (CVD). We observe a transient decrease of THz absorption in graphene subject to pulsed optical excitation, a result in contrast with the increased absorption reported previously in epitaxial graphene.\cite{George2008,Choi2009,Strait2011}  In addition, the differential THz conductivity spectrum deviates significantly from the Drude form.  We propose that the observed anomalous THz bleaching arises from additional scattering of electrons with optical phonons in graphene and on the substrate, as well as thermal broadening of the electron distribution under non-equilibrium conditions.

The apparatus and procedure used in our optical-pump THz-probe experiment are similar to those used in Ref (\onlinecite{Beard2001}). A regeneratively amplified Ti:Sapphire system provided pulses of central wavelength 800 nm, pulse duration $\sim$100 fs and repetition rate 5 kHz. The THz probe pulses were generated by optical rectification of the 800 nm pulses in a 1 mm-thick ZnTe crystal and detected by electro-optic sampling of the 800 nm pulses in a 0.1 mm-thick ZnTe crystal. We used a thin detection crystal to minimize errors associated with electro-optic sampling, which become significant when the sample response varies on the time scale of the THz pulse ($\sim$ 1 ps, as was the case in our time-resolved measurements).\cite{Beard2001} For equilibrium measurements, the THz signal was measured using balanced photodiodes and a lock-in amplifier phase-locked to a mechanical chopper, which modulated the THz generation beam. For time-resolved non-equilibrium measurements, the pump beam was chopped to allow detection of the small differential changes in the signal induced by the pump pulses. The spot sizes of the 800 nm pump beam and the THz probe beam on the sample were $\sim$5 mm and $\sim$2 mm, respectively, to ensure approximately homogeneous photoexcitation of the probed sample area. Measurements presented here were performed in the low-fluence regime (incident pump fluence $\sim$10 $\mu$J/cm$^{2}$) at room temperature and in high vacuum (pressure $< 10^{-5}$ Torr). Using the two-temperature model of Ref (\onlinecite{Lui2010}), we estimate that under our experimental conditions, the maximum transient electronic temperature was $\lesssim$ 1000 K and the full equilibrium heating of the lattice was negligible.  

Monolayer graphene samples were grown by CVD\cite{Li2009} on  copper foils and subsequently transferred to different substrates, including sapphire, $z$-cut crystalline quartz and borosilicate glass.  The monolayer thickness and the sample quality of the CVD graphene were characterized by Raman spectroscopy\cite{Ferrari2006}   (inset of Fig. 1). 
\begin{figure}
   \includegraphics{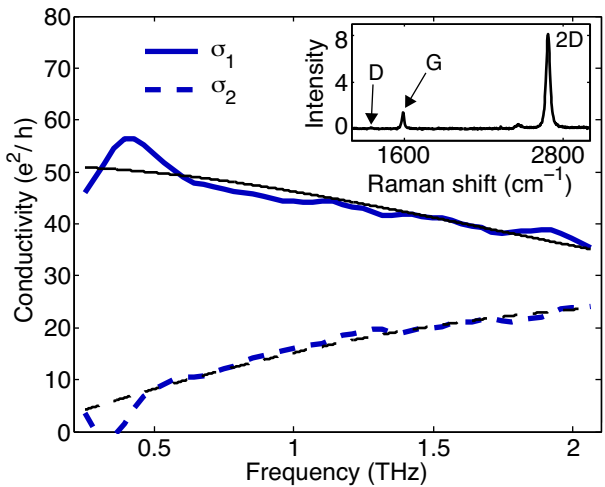}
   \caption{\label{charact} (color) Complex optical sheet conductivity of graphene from 0.2 to 2.1 THz, in units of $e^2/h$.  The measurement was performed at room temperature on monolayer CVD graphene on a quartz substrate (without optical excitation). A bare quartz substrate was used as a reference.  The blue solid and dashed lines are the real ($\sigma_1$) and imaginary ($\sigma_2$) part of the conductivity, respectively. The data can be fit by the Drude model (thin black lines) with carrier scattering rate $\Gamma$ = 3 THz. The inset shows a typical Raman spectrum of our CVD graphene (excitation wavelength 532 nm).  The narrow Lorentzian line shape of the 2D mode confirms the monolayer thickness of the samples.  The small D-mode signal indicates the high crystalline quality of the samples.}
\end{figure}
We have measured the THz absorption spectrum of the CVD samples without optical excitation and extracted the complex optical conductivity of graphene using the method described in Ref (\onlinecite{Nuss1998}). Fig. 1 displays the conductivity spectrum of a graphene sample on a quartz substrate at room temperature.  The data can be fit well by the Drude formula with a Fermi energy $E_{\rm F} \sim$300 meV and scattering rate $\Gamma\sim3$ THz (100 cm$^{-1}$), corresponding to a carrier density $n$ $\sim 6 \times 10^{12}$ cm$^{-2}$ and mobility $\mu$ = 2000 cm$^2$/V$\cdot$s.  These parameters are reasonable for doped CVD graphene on a substrate.\cite{Yan2011,Lee2011,Horng2011,Ren2012}

When we pump the graphene/quartz sample with 800 nm pulses, we observe a significant change in the transmitted THz probe pulses (Fig. 2a). Strikingly, the THz transmission is found to increase following pulsed excitation.  For a thin film deposited on a transparent substrate, the differential transmitted electric field $\Delta E$, normalized to the equilibrium transmitted field $E$, is related to the differential optical conductivity $\Delta \sigma$ as\cite{Nuss1998}
\begin{equation}
\frac{\Delta E}{E} = -\left(\frac{Z_0}{n_s + 1}\right)\Delta\sigma
\end{equation}
where $n_s$ is the substrate refractive index and $Z_0$ the impedance of free space.  Our observation of positive $\Delta E$ therefore corresponds to negative differential conductivity, or reduced absorption, in graphene.  Fig. 2b displays the temporal dynamics of the pump-induced modulation of the THz transmission, which are well described by an exponential decay with time constant $\tau = 1.7$ ps (Fig. 2b inset).  It is firmly established that photoexcited charge carriers in graphene thermalize rapidly and relax most of their energy to a set of strongly coupled optical phonons within a few hundred femtoseconds.\cite{Kampfrath2005,Lui2010,Butscher2007,Yan2009,Kang2010,Wu2012}  The equilibrated subsystem of electrons and optical phonons subsequently cools within a few picoseconds through the anharmonic decay of the optical phonons.  We therefore attribute the observed dynamics of the THz response to the cooling of the coupled electron-phonon system. 
\begin{figure}
   \includegraphics{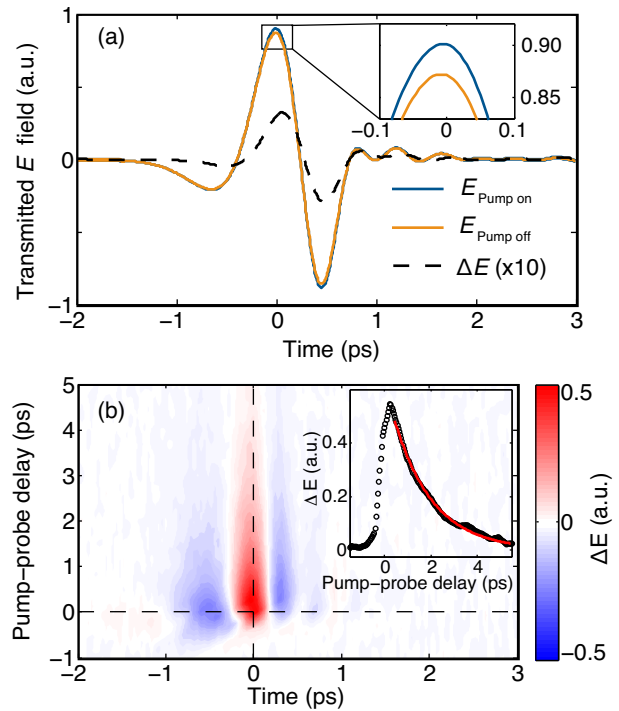}
   \caption{\label{pumponoff} (color) (a) Transmission of the THz electric field with (blue) and without (orange) pump excitation, measured by chopping the THz generation beam.  The dashed line, scaled by a factor of 10 for clarity, is the pump induced modulation of the transmitted electric field $\Delta E$.  The inset shows a zoomed-in view of the peak, indicating that $\Delta E$ corresponds to a pump-induced bleaching of the graphene sample.  (b) Temporal dynamics of $\Delta E$ following optical pump excitation, measured by chopping the pump beam.  The horizontal and vertical dashed lines are, respectively, the zero pump-probe delay time and the peak position of $\Delta E$.  Inset shows the temporal dynamics of the peak of $\Delta E$ (vertical dashed line in main panel). The red line is a single-exponential fit with time constant $\tau = 1.7$ ps. The peak value of the signal corresponds to $\Delta E/E\sim 5\%$.}
\end{figure}

We have used Eq. (1) to extract the complex differential conductivity spectra ($\Delta \sigma = \Delta\sigma_1 + i \Delta \sigma_2$) from the transmission data in Fig. 2b at different pump-probe delay times (Fig. 3a). We find that $\Delta\sigma_1$ remains negative for the whole decay process after pulsed excitation (Fig. 3a), and for the entire measured spectral range (see, for example, the spectrum at pump-probe delay 1 ps in Fig. 3b).  We note that we have observed negative $\Delta\sigma$ of similar magnitude and lifetime (both with variation $<20$\%) at temperatures ranging from 4 to 300 K. The response is also found to be similar for CVD graphene samples on different substrates (sapphire, quartz and borosilicate glass), and in both ambient and vacuum conditions. We therefore conclude that it is a general property of highly doped graphene on a substrate. The results are surprising because the intraband absorption of graphene is typically described by the Drude model with a constant scattering rate. Increasing the free carrier population by photoexcitation should lead to enhanced THz absorption, as observed in epitaxial graphene layers on SiC substrate,\cite{George2008,Choi2009,Strait2011} as well as in traditional semiconductors such as GaAs\cite{Beard2001} and Si.\cite{Suzuki2009} The explanation of our experimental data must therefore lie beyond this simple picture.

\begin{figure}
   \includegraphics{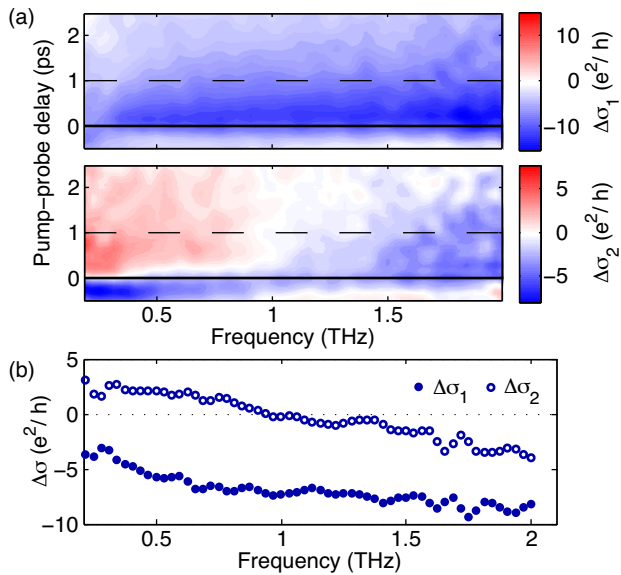}
   \caption{\label{photocon} (color) (a) Temporal dynamics of the real (upper panel) and imaginary (lower panel) parts of the differential THz conductivity of optically pumped monolayer graphene. The solid black line denotes zero pump-probe delay time.  (b) Differential THz conductivity 1 ps after optical excitation (horizontal dashed lines in (a)).  The filled and open circles denote the experimental real ($\Delta\sigma_1$) and imaginary ($\Delta\sigma_2$) parts of the conductivity, respectively. }
\end{figure}

The negative sign and non-Drude spectral shape of the measured differential THz conductivity in graphene (Fig. 3b) can be qualitatively understood by considering the increased scattering rate and broadened carrier distribution present in the transient regime, where the electron and phonon systems are driven to considerably higher temperatures than in equilibrium. After pulsed excitation, thermalization and cooling of photoexcited carriers are expected to proceed via emission of strongly coupled optical phonons\cite{Yan2009, Kang2010, Wu2012} and remote substrate phonons.\cite{Low2012} The enlarged phase space for scattering at higher carrier temperature, combined with the increased phonon populations, can result in significantly stronger electron-phonon scattering. The suppression of free-carrier conductivity due to electron-phonon scattering at elevated temperature has been observed in ultrafast studies of graphite\cite{Kampfrath2005} and DC transport studies of graphene.\cite{Chen2008} This mechanism is also responsible for current saturation in high-field transport studies of graphene devices.\cite{Barreiro2009,Chae2010,Berciaud2010,Perebeinos2010}   A recent THz-pump THz-probe study also showed signatures of increased intraband scattering due to heating the electron system with an intense THz pulse.\cite{Hwang2011} We therefore expect that increased electron-phonon scattering in the transient regime will lead to the observed negative differential THz conductivity. Additionally, the energy dependence of carrier scattering must be considered because of the thermally broadened distribution of the hot ($k_{\rm B}T \sim \mu$) carriers.\cite{Tielrooij2012} This will lead to the observed non-Drude spectral shape because the Drude model assumes a constant (energy independent) scattering rate.\cite{Ashcroft1976} More generally, to achieve a full quantitative understanding of our data, other factors should be considered. Recent research has provided evidence that there is a transient regime of population inversion in graphene during the first few picoseconds after optical pump excitation,\cite{Li2012} which could amplify a THz probe pulse by stimulated emission.\cite{Otsuji2012}  Such effects are not ruled out by our data. Further studies at higher fluence and on samples with different Fermi energies or layer thicknesses will be needed to clarify the roles played by each of these mechanisms.

In summary, we have observed reduced absorption of THz radiation and a non-Drude differential conductivity spectrum in graphene subject to pulsed optical excitation. Our results can be explained by additional electron-phonon scattering in conjunction with a thermally broadened carrier distribution.  This work demonstrates that the THz response of graphene is strongly tunable by optical means over a broad frequency range on an ultrafast picosecond timescale. 

Upon completion of this work, we became aware of similar works by other groups.\cite{Docherty2012,Jnawali2013}  As mentioned in the main text, however, we did not observe any influence of the ambient environment on the sign of the differential conductivity, as reported therein.\cite{Docherty2012}

The authors acknowledge helpful discussions with N.M. Gabor, J.C.W. Song, and H.Y. Hwang. This research is supported by Department of Energy Office of Basic Energy Sciences Grant No DE-SC0006423. A.J.F.acknowledges support from NSF GRFP.P.J.H acknowledges support by the Air Force Office of Scientific Research and the Office of Naval Research GATE MURI. This work made use of MIT's MRSEC Shared Experimental Facilities supported by the National Science Foundation under award No. DMR-0819762 and of HarvardÕs Center for Nanoscale Systems (CNS), supported by the National Science Foundation under grant No. ECS-0335765.

\end{document}